\documentclass[preprintnumbers,amsmath,amssymb,superscriptaddress,pra,twocolumn]{revtex4-2}

\include{amsmath}
\usepackage{graphicx}
\usepackage{subfigure}
\usepackage{amsmath}
\usepackage{amsfonts}
\usepackage{color}
\usepackage[latin1]{inputenc}
\usepackage{bbold}
\usepackage{times}

\begin{document}

\title[]{Quantum state purity versus average phonon number for characterization \\  of mechanical oscillators in cavity optomechanics\\
}
\author{K. B{\o}rkje}
  \affiliation{Department of Science and Industry Systems, University of
  South-Eastern Norway, NO-3603 Kongsberg, Norway}
\author{F. Marin}
\affiliation{Dipartimento di Fisica e Astronomia, Università di Firenze, I-50019 Sesto Fiorentino (FI), Italy} \affiliation{European Laboratory for Non-Linear Spectroscopy (LENS), I-50019 Sesto Fiorentino (FI), Italy} \affiliation{INFN, Sezione di Firenze, I-50019 Sesto Fiorentino (FI), Italy}
\affiliation{CNR-INO, I-50125 Firenze, Italy}
\date{\today}


\begin{abstract}
Quantum oscillators in Gaussian states are often characterized by average occupation numbers that refer to a basis of eigenstates of the non-interacting oscillator Hamiltonian. We argue that quantum state purity is a more appropriate characteristic of such states, which can be applied to oscillators of any dimensionality. For a one-dimensional oscillator, the state purity is directly related to a thermal occupation number defined with respect to the number state basis in which the oscillator's quantum state is thermal. Thus, it naturally introduces a more versatile definition of an average occupation number. We study optomechanical sideband cooling of one- and two-dimensional mechanical oscillators in particular, and derive exact analytical expressions for the maximal mechanical state purity achievable in the quantum backaction limit. In the case of a one-dimensional oscillator, we show that the thermal occupation number related to purity can be well approximated by the average phonon number in the weak-coupling regime, but that the two differ in the regime of ultrastrong optomechanical coupling or in cases where the oscillator's resonance frequency is strongly renormalized.
\end{abstract}

\maketitle

\twocolumngrid

\section{Introduction} 

The quantum-mechanical nature of mechanical oscillators has been probed in a wide variety of experiments, ranging from the microscopic scale with the motion of trapped ions \cite{Wineland2013RMP} to the meso- or macroscopic scale with, e.g., flexural modes of silicon nitride \cite{Purdy2015PRA,Underwood2015PRA} or aluminum \cite{Teufel2011Nature} membranes, bulk acoustic-wave resonators in solids \cite{OConnell2010Nature,Chu2018Nature} or liquids \cite{Shkarin2019PRL}, or the motion of optically levitated nanoparticles \cite{Delic2020Science,Magrini2021Nature,Tebbenjohanns2021Nature,Ranfagni2022PRR,Piotrowski2022}.

In the absence of nonlinearities, i.e., when oscillators are limited to Gaussian states, genuine signatures of nonclassicality are lacking. It is then common to assess the quantum nature of the oscillator quantitatively. In particular, in cooling experiments where the goal is to remove energy from a one-dimensional harmonic mechanical oscillator, either passively through interaction with a cold reservoir or through active feedback, it is customary to characterize the oscillator by its average phonon number. This is the average number of excitations in the basis of eigenstates of the {\it isolated} harmonic oscillator Hamiltonian, which we will refer to as the phonon basis, and is defined with reference to the oscillator's bare resonance frequency. A focus on achieving the lowest possible average phonon number may then suggest that the lower the number, the ``more quantum" the oscillator is.

The reliance on average phonon number as a measure of quantumness of an oscillator is, however, problematic for a number of reasons. 
One reason is that if the cooling mechanism affects position and momentum fluctuations differently, the representation of the state of the oscillator in the phonon basis may not be thermal. Characterizing the state by its average phonon number means, e.g., that the squeezed vacuum state will be deemed ``less quantum" than the ground state even though it is arguably ``more quantum" in the particular sense that a true probability distribution of the Glauber-Sudarshan type does not exist \cite{Yuen1976PRA}. Another reason is that the mechanical system might bear little resemblence to a harmonic oscillator at the bare resonance frequency. The interactions may have caused the oscillator to hybridize with other degrees of freedom such that the mechanical system can no longer be viewed as a simple harmonic oscillator at all and that its spectral weight is distributed at frequencies far from the bare resonance frequency of the isolated system. A third problem with the average phonon number is that it is not obvious how to generalize it in order to define a single number for characterizing higher-dimensional mechanical oscillators. 



In this article, we argue that for quantifying the quantum character of mechanical oscillators in Gaussian states, quantum state purity
\begin{equation}
\label{eq:Purity}
\mu = \mathrm{Tr}\left(\hat{\rho}^2\right) ,
\end{equation}
where $\hat{\rho}$ is the density matrix representing the state, is a more appropriate measure than average phonon number. For any oscillator dimensionality, the purity of a Gaussian state can be calculated from expectation values of quadratic functions of position and momentum fluctuations without having to refer to resonance frequencies associated with a confining potential. Furthermore, we show that for one-dimensional oscillators in Gaussian states, the concept of purity leads to a natural definition of a thermal occupation number which fully determines how much the purity deviates from unity. This thermal occupation number is simply the average number of excitations in the number state basis in which the state $\hat{\rho}$ is thermal, which always exists. It generally differs from the average phonon number but coincides with it for a harmonic oscillator at the bare resonance frequency in a thermal state. 

While these general considerations can apply to a variety of systems, we study optomechanical sideband cooling in particular in this article. We start by modeling a two-dimensional oscillator coupled linearly to a single cavity mode. By the introduction of bright and dark mechanical modes, we show under which special circumstances the problem reduces to the canonical optomechanical system of a single mechanical mode coupled to a single cavity mode. This special case is studied first, and the theory then applies to a wide variety of experimental realizations of cavity optomechanics \cite{Aspelmeyer2014RMP}. We subsequently proceed to the setup where the cavity mode couples to the full two-dimensional motion of the mechanical system, with particular relevance to recent experiments with levitated nanoparticles \cite{Ranfagni2022PRR,Piotrowski2022}.

In optomechanics, the thermal occupation number associated with mechanical state purity has, to our knowledge, previously only been discussed in the context of mechanical squeezing \cite{Kronwald2013PRA_2}. Here, we derive the thermal occupation number for optomechanical sideband cooling of a one-dimensional oscillator -- a case for which it was recently measured in an experiment \cite{Ranfagni2022PRR}. We first show that it matches the average phonon number in the regime of weak optomechanical coupling, as long as the phonon number is defined according to the {\it effective} resonance frequency which includes a shift due to the optical spring effect. We then show that in the regime of ultrastrong optomechanical coupling \cite{Peterson2019PRL,Sommer2021NatComm,Ranfagni2021NatPhys,Ranfagni2022PRR}, where the (laser-drive enhanced) coupling rate becomes comparable to the bare oscillator resonance frequency, the thermal occupation number no longer matches the phonon number. For the problem with a two-dimensional oscillator, we calculate the purity of the oscillator's state and show under which circumstances it can come close to unity.

For optomechanical sideband cooling, the theoretical upper limit for the mechanical state purity can be determined by ignoring all noise sources except the electromagnetic vacuum noise entering the cavity mode. This is the so-called {\it quantum backaction limit}. We present exact analytical expression for the purity in this limit, both for one- and two-dimensional oscillators. 

The article is organized as follows. In Section \ref{sec:Purity}, we show how quantum state purity relates to observable expectation values for one- and two-dimensional oscillators in Gaussian states, and we define the related thermal occupation number in the one-dimensional case. We define the optomechanical model in Section \ref{sec:Model}. In Sections \ref{sec:1DOM} and \ref{sec:2DOM}, we calculate the purity for optomechanical systems with one- and two-dimensional oscillators, respectively, compare with the standard average phonon number, and present analytical expressions for the upper limits of purity. We conclude in Section \ref{sec:Conclusion}. For convenience, an overview of the symbols used in the article is found in the Appendix. 

\section{Purity and thermal occupation of oscillators in Gaussian states}
\label{sec:Purity}

In this section, we present how quantum state purity of 1D and 2D oscillators in Gaussian states can be expressed in terms of observable expectation values of positions and momenta. We also discuss how the state purity measure leads to a natural definition of average thermal occupation numbers in the case of Gaussian states.

\subsection{One-dimensional oscillator}

Let us consider a general one-dimensional oscillator with position and momentum operators $\hat{x}$ and $\hat{p}$ satisfying the canonical commutation relation
\begin{equation}
\label{eq:CommRel}
\left[\hat{x} , \hat{p} \right] = i \hbar .
\end{equation}
The system's quantum state $\hat{\rho}$ is assumed to be Gaussian, meaning that the corresponding Wigner quasiprobability distribution is a Gaussian. The state can then always be expressed as \cite{Adam1995JModOpt,Paris2003PRA}
\begin{equation}
\label{eq:rho}
\hat{\rho} =  \sum_{n=0}^\infty \frac{\bar{n}^n}{\left(\bar{n} + 1\right)^{n+1}} |n \rangle \langle n | ,
\end{equation}
i.e., there always exists a number state basis in which $\hat{\rho}$ is a thermal state. We define associated bosonic creation and annihilation operators $\hat{b}^\dagger, \hat{b}$, where $[ \hat{b},\hat{b}^\dagger ]=  1$ and $\hat{b}|n\rangle = \sqrt{n} \, |n-1\rangle $. It follows that
\begin{align}
\label{eq:nmFromBs}
\bar{n} = \mathrm{Tr} \left(\hat{b}^\dagger \hat{b} \hat{\rho} \right) .
\end{align}
Since $\hat{\rho}$ is diagonal in the number state basis $|n\rangle$, we can interpret it as a classical probabilistic mixture of states with differing excitation numbers $n$ or, in a particle interpretation, differing number of particles. The number $\bar{n}$ is then interpreted as the average number of excitations or, equivalently, the average particle number. 

The purity of the state $\hat{\rho}$ straightforwardly follows from the orthonormality of the basis states $|n\rangle$, giving
\begin{equation}
\label{eq:PurityThermal}
\mu =  \frac{1}{2 \bar{n} + 1} .
\end{equation}
The limit of a pure state $\hat{\rho}$ thus corresponds to $\bar{n} \rightarrow 0$, i.e., when $\hat{\rho} = |0\rangle \langle 0|$. 

We now define relations between the creation and annihilation operators $\hat{b}^\dagger, \hat{b}$ and the position and momentum operators by
\begin{align}
\label{eq:zpfDef1}
  \hat{x}  & = \bar{x} +  x_\mathrm{zpf} \left(\hat{b} + \hat{b}^\dagger \right) \\
\label{eq:zpfDef2} \hat{p}  & = \bar{p} + i p_\mathrm{zpf} \left(e^{-i \theta } \hat{b}^\dagger -  e^{i \theta }\hat{b} \right) .
\end{align}
The parameters $\bar{x}, \bar{p}$ must then satisfy $\bar{x} = \langle \hat{x} \rangle$ and $\bar{p} = \langle \hat{p} \rangle $. The remaining four parameters $x_\mathrm{zpf}$, $p_\mathrm{zpf}$, $\bar{n}$, and $\theta$ must be chosen so as to satisfy the commutation relation \eqref{eq:CommRel} and give correct values for the three distinct elements of the covariance matrix, $\langle \Delta \hat{x}^2 \rangle$, $\langle \Delta \hat{p}^2 \rangle$, $\langle \{ \Delta \hat{x},\Delta \hat{p} \} \rangle/2$ (which completely specifies a Gaussian state), when defining
\begin{align}
\label{eq:FluctDef}
\Delta \hat{o} = \hat{o} - \langle \hat{o} \rangle .
\end{align}
From this, it follows that the average thermal occupation number, as defined by Equation \eqref{eq:rho}, is 
\begin{align}
\label{eq:BOccNum2}
2 \bar{n} + 1 = \frac{1}{\hbar} \sqrt{4 \langle \Delta \hat{x}^2 \rangle \langle \Delta \hat{p}^2 \rangle -  \langle \left\{ \Delta \hat{x} ,  \Delta \hat{p} \right\}\rangle^2 } .
\end{align}
The purity of the state $\hat{\rho}$ in Equation \eqref{eq:PurityThermal} is in other words inversely proportional to the square root of the determinant of the covariance matrix \cite{Paris2003PRA}. Equation \eqref{eq:BOccNum2} thus allows for obtaining the thermal occupation number, and thereby the purity, solely from observations of position and momentum fluctuations. We note that beyond the paradigm of Gaussian states, a quantity $\bar{n}$ {\it defined} according to Equation \eqref{eq:BOccNum2} is a measure of deviation from a minimal uncertainty state \cite{Ranfagni2022PRR}.

Furthermore, one finds that $\theta$, $x_\mathrm{zpf}$, and $p_\mathrm{zpf}$ must be
\begin{align}
\label{eq:anglesin}
\sin \theta = \frac{\langle \{\Delta \hat{x} , \Delta \hat{p}\} \rangle}{2\sqrt{\langle \Delta \hat{x}^2 \rangle  \langle \Delta \hat{p}^2 \rangle}} ,
\end{align}
\begin{align}
\label{eq:xzpf}
x^2_\mathrm{zpf} & = \frac{\hbar \langle \Delta \hat{x}^2 \rangle }{\sqrt{4 \langle \Delta \hat{x}^2 \rangle  \langle \Delta \hat{p}^2 \rangle - \langle \{\Delta \hat{x} , \Delta \hat{p}\} \rangle^2}} ,
\end{align}
and
\begin{align}
\label{eq:pzpf}
p^2_\mathrm{zpf} & = \frac{\hbar \langle \Delta \hat{p}^2 \rangle }{\sqrt{4 \langle \Delta \hat{x}^2 \rangle  \langle \Delta \hat{p}^2 \rangle - \langle \{\Delta \hat{x} , \Delta \hat{p}\} \rangle^2}} .
\end{align}
We note that since $\langle \Delta \hat{x}^2  \rangle = x_\mathrm{zpf}^2 (2 \bar{n} + 1)$ ($\langle \Delta \hat{p}^2  \rangle = p_\mathrm{zpf}^2 (2 \bar{n} + 1)$), one can think of the parameter $x_\mathrm{zpf}$ ($p_\mathrm{zpf}$) as the size of the zero-point fluctuations in position (momentum) in the state $|0\rangle \langle 0|$, i.e., the ground state of the basis in which $\hat{\rho}$ is a thermal state. 

The above discussion shows that one can always consider a one-dimensional system whose reduced state $\hat{\rho}$ is Gaussian as being in a probabilistic mixture of pure states $|n\rangle$. This is true even in the presence of interactions with other systems. From the relations \eqref{eq:zpfDef1}, \eqref{eq:zpfDef2}, one finds that the position space representations of the basis states $|n\rangle$ are $\Psi_n(x) = \langle x |n\rangle = \psi_n(x-\bar{x}) e^{i\bar{p}(x-\bar{x})/\hbar}$, where
\begin{align}
\label{eq:Psin}
\psi_n(x) & = \frac{1}{\sqrt{2^n n!}} \left(\frac{|M \Omega|}{\pi \hbar}\right)^{1/4} H_n\left(\sqrt{\frac{\mathrm{Re} \, M \Omega}{\hbar}}x \right) e^{- \frac{M \Omega}{2\hbar} x^2}  ,
\end{align}
having defined
\begin{align}
\label{eq:OmegaComplexTildelambdaDef}
M \Omega & = e^{-i\theta} \frac{p_\mathrm{zpf}}{x_\mathrm{zpf}} , 
\end{align}
and where $H_n$ are the Hermite polynomials. The wavefunctions $\psi_n$, sometimes referred to as generalized harmonic-oscillator states in the literature \cite{Meyer1981ChemPhys,Moller1996PRA}, have the form of the standard energy eigenfunctions for an isolated harmonic oscillator with mass $M$ and frequency $\Omega/(2\pi)$. However, $M \Omega$ is here a complex parameter defined by the entries of the covariance matrix. 

Finally, we note that when transforming to the phonon basis, the pure state $|0 \rangle \langle 0 |$ defined above will correspond to either the vacuum, a coherent state, a squeezed vacuum state, or a squeezed coherent state. The reason, as will be shown below (see Equation \eqref{eq:AnnihilationRelation}), is that the transformation can be decomposed into squeezing and displacement transformations. 

\subsection{Two-dimensional oscillator}
\label{sec:2DGauss}

We now consider a two-dimensional oscillator whose position and momentum operators have components $(\hat{x}, \hat{y})$ and $(\hat{p}_x,\hat{p}_y)$, respectively, satisfying 
\begin{align}
\label{eq:CommRel2D}
\left[\hat{x} , \hat{p}_x \right] = \left[\hat{y} , \hat{p}_y \right] & = i \hbar \\
\left[\hat{x} , \hat{p}_y \right] = \left[\hat{y} , \hat{p}_x \right] & = 0 .
\end{align}
If the oscillator is in a Gaussian state, it can be written as a tensor product of one-mode thermal states subjected to a unitary transformation \cite{Serafini2004JPhysB}. This means that we may write
\begin{equation}
\label{eq:GenGaussStateTwoModes}
\hat{\rho} = \sum_{m = 0}^\infty \sum_{n = 0}^\infty \frac{\bar{m}^m \bar{n}^n}{\left(\bar{m} + 1\right)^{m+1}\left(\bar{n} + 1\right)^{n+1}} |\Psi_{m,n} \rangle \langle \Psi_{m,n}|
\end{equation}
where $|\Psi_{m,n}\rangle$ is an orthonormal basis for the two-mode system, i.e., $\langle \Psi_{m,n}|\Psi_{m',n'}\rangle = \delta_{m,m'} \delta_{n,n'}$, and $\bar{m}, \bar{n}$ are thermal occupation numbers. In this case, the purity of the state $\hat{\rho}$ can be expressed as
\begin{equation}
\label{eq:PurityGaussStateTwoModes}
\mu_\mathrm{2D} = \frac{1}{(2 \bar{m} + 1)(2 \bar{n} + 1) } .
\end{equation}

The purity is again proportional to the inverse square root of the covariance matrix \cite{Serafini2004JPhysB}, just as in the one-dimensional case. For simplicity, we will now assume  
\begin{equation}
\label{eq:xpnotcorr}
\langle \{\Delta \hat{x}, \Delta \hat{p}_x\} \rangle = \langle \{\Delta \hat{y},\Delta \hat{p}_y\} \rangle  =  0 
\end{equation}
and
\begin{equation}
\label{eq:xpCrossOpp}
\langle \Delta \hat{y} \Delta \hat{p}_x \rangle = - \langle \Delta \hat{x} \Delta \hat{p}_y \rangle ,
\end{equation}
which are satisfied in the model we will study below. This gives the relation
\begin{align}
\label{eq:Purity2D}
\mu_\mathrm{2D} & = \frac{\left(\hbar/2\right)^2}{\sqrt{A_{xx} A_{pp} - A_{xp} B_{xp} + B_{xp}^2}}
\end{align}
between state purity and the (in principle) observable expectation values of position and momentum fluctuations, where we have defined
\begin{align}
\label{eq:Purity2DHelp}
A_{xx} & = \langle \Delta \hat{x}^2 \rangle \langle  \Delta \hat{y}^2 \rangle - \langle  \Delta \hat{x}  \Delta \hat{y} \rangle^2 \\
A_{pp} & = \langle  \Delta \hat{p}_x^2 \rangle \langle  \Delta \hat{p}_y^2 \rangle - \langle  \Delta \hat{p}_x  \Delta \hat{p}_y \rangle^2 \\
A_{xp} & = \langle  \Delta \hat{x}^2 \rangle  \langle  \Delta \hat{p}_y^2 \rangle +  \langle  \Delta \hat{y}^2 \rangle  \langle  \Delta \hat{p}_x^2 \rangle \notag \\
& - 2 \langle  \Delta \hat{x}  \Delta \hat{y} \rangle \langle  \Delta \hat{p}_x  \Delta \hat{p}_y \rangle \\
B_{xp} & =  \langle  \Delta \hat{x}  \Delta \hat{p}_y \rangle^2 .
\end{align}
We note that when \eqref{eq:xpnotcorr} and \eqref{eq:xpCrossOpp} are fulfilled for one choice of orthogonal coordinates $\hat{x}, \hat{y}$, they are also valid for any other choice of orthogonal coordinates.

\section{Model}
\label{sec:Model}

Consider now a two-dimensional mechanical oscillator with mass $m$ whose motion is coupled to a single cavity mode's field fluctuations. The oscillator's position operator has components $\hat{x}$ and $\hat{y}$, with respective canonically conjugate momentum operators $\hat{p}_x$, $\hat{p}_y$. We choose the $x$- and $y$-axis as the principal axes of the harmonic trap which defines the oscillator, with $\omega_x/(2\pi)$, $\omega_y/(2\pi)$ the associated resonance frequencies. 

\begin{figure}[htb]
\includegraphics[width=.99\columnwidth]{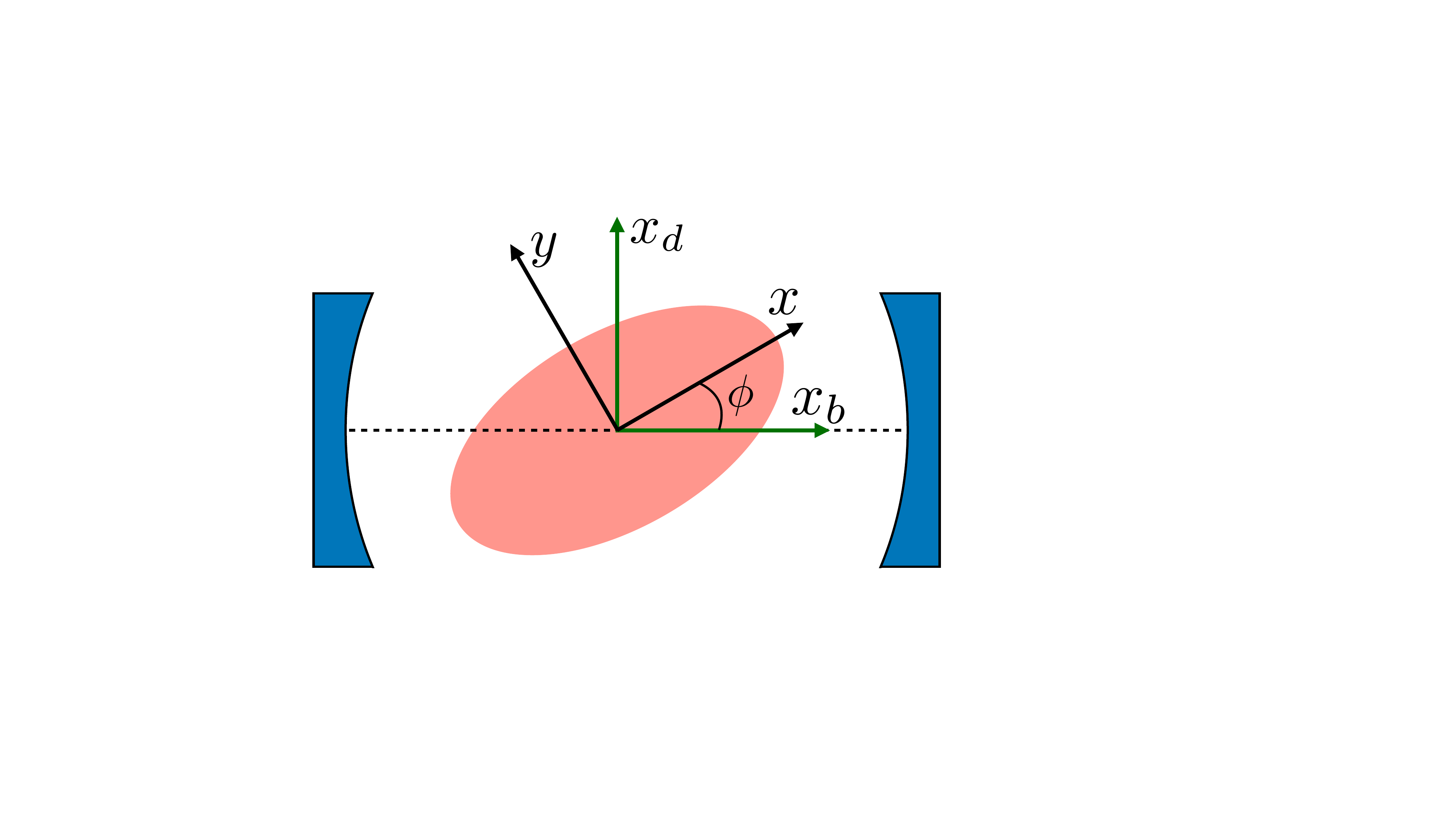}
\caption{Setup. The axes $(x,y)$ are the principal axes of the harmonic potential defining the two-dimensional mechanical oscillator. An alternative coordinate system with orthogonal axes $(x_b,x_d)$ is also shown. Only the motion along the $x_b$-axis, i.e., the cavity axis, couples directly to the cavity mode. For this reason, we refer to $x_b$ as the {\it bright mode} direction and $x_d$ as the {\it dark mode} direction.}
\label{fig:Setup}
\end{figure}

The model we study can apply to two different scenarios: (A) a nanoparticle levitated by an optical tweezer and coupled to the cavity mode by coherent scattering \cite{Delic2019PRL,Windey2019PRL,Toros2020PRR,Toros2021PRR} and (B) a two-dimensional oscillator, either clamped \cite{Gloppe2014NatNano} or levitated \cite{Kiesel2013ProcNatlAcadSciUSA}, whose coupling to the cavity mode comes about due to direct laser driving of the cavity mode. In both scenarios, we assume that the cavity mode only couples to the component of motion along the cavity axis. This axis is rotated by the angle $\phi$ from the $x$-axis, such that the cavity mode couples to the linear combination
\begin{align}
\label{eq:BrightDef}
\hat{x}_b = \cos \phi \, \hat{x} - \sin \phi  \, \hat{y} .
\end{align}
For this reason, it will be convenient below to express the model in terms of a {\it bright mode} with associated position operator $\hat{x}_b$ and an orthogonal {\it dark mode}, which does not directly couple to the cavity mode, with position operator
\begin{align}
\label{eq:DarkDef}
\hat{x}_d = \sin \phi \, \hat{x} + \cos \phi  \, \hat{y} .
\end{align}
We define canonically conjugate momentum operators $\hat{p}_b$, $\hat{p}_d$ for the bright and dark modes accordingly. The two orthogonal coordinate systems $(x,y)$ and $(x_b,x_d)$ are illustrated in Figure \ref{fig:Setup}.

The system is described by the Hamiltonian
\begin{align}
\label{eq:Hamiltonian}
H & = \frac{\hat{p}_x^2}{2m} + \frac{1}{2} m \omega_x^2 \hat{x}^2  +  \frac{\hat{p}_y^2}{2m} + \frac{1}{2} m \omega_y^2 \hat{y}^2 \\
& + \hbar \Delta \hat{a}^\dagger \hat{a} + \hbar \lambda_o \left(\cos \phi \, \hat{x} - \sin \phi  \, \hat{y} \right) \left(\hat{a} + \hat{a}^\dagger \right) \notag .
\end{align}
In the levitated nanoparticle scenario (A), the quantity $-\Delta/(2\pi)$ is the detuning between the optical tweezer frequency and the resonance frequency of the cavity mode $\omega_c$, and $\hat{a}$ is the photon annihilation operator for the cavity mode \cite{Delic2019PRL,Windey2019PRL}. In the alternative scenario with direct cavity driving (B), the cavity mode is coherently driven at a frequency detuned by $-\Delta/(2\pi)$ from its resonance frequency $\omega_c$, in which case $\hat{a}$ is the {\it displaced} photon annihilation operator (see e.g.~\cite{Aspelmeyer2014RMP}) describing fluctuations relative to a coherent state for the cavity mode. The interaction between mechanical and cavity fluctuations, which scales with the amplitude of the tweezer field in scenario (A) or with the coherent drive amplitude in scenario (B), is quantified by a parameter $\lambda_o$ which we define positive without loss of generality. For the levitated nanoparticle setup (A), we note that $\lambda_o \propto \cos \phi$ when $y$ is the direction of linear polarization of the optical tweezer \cite{Delic2019PRL,Windey2019PRL}, but for a fixed $\phi \neq \pi/2$, we may view it simply as a constant in the following. 

We now move to a description in terms of quantum Langevin equations in order to include coupling to the mechanical and cavity modes' external baths \cite{Gardiner1985PRA,Giovannetti2001PRA}. We will assume that the motion in the $x$- and $y$-directions are subject to independent Brownian quantum noise. This gives the equations \cite{Giovannetti2001PRA}
\begin{align}
\label{eq:QLE0}
\dot{\hat{x}} & = \frac{\hat{p}_x}{m} \\
\dot{\hat{p}}_x & = -\gamma_x \hat{p}_x - m\omega_x^2 \hat{x} +  \hat{N}_x  - \hbar \lambda_o \cos \phi \left(\hat{a} + \hat{a}^\dagger \right)  \\
\dot{\hat{y}} & = \frac{\hat{p}_y}{m} \\
\dot{\hat{p}}_y & = -\gamma_y \hat{p}_y - m\omega_y^2 \hat{y} +  \hat{N}_y  + \hbar \lambda_o \sin \phi \left(\hat{a} + \hat{a}^\dagger \right)   
 \end{align}
describing the mechanical oscillator. We have introduced $\gamma_x$ and $\gamma_y$ as the bare energy decay rates of the mechanical modes. Defining the Fourier transformation according to 
\begin{align}
\label{eq:FourDef}
f^{(\dagger)}[\omega] = \int_{-\infty}^\infty dt \, e^{i\omega t} f^{(\dagger)}(t) ,
 \end{align}
the Gaussian mechanical Brownian noise operators $\hat{N}_x$, $\hat{N}_y$ are uncorrelated and both satisfy the relation \cite{Giovannetti2001PRA}
\begin{align}
\label{eq:NoiseRel}
\langle \hat{N}_j[\omega] \hat{N}_j[\omega'] \rangle  & = \hbar m \gamma_j \omega \left[\coth\left(\frac{\hbar\omega}{2 k_B T}\right) + 1 \right] \\
& \times 2\pi \delta(\omega + \omega') , \notag 
 \end{align}
where $k_B$ is the Boltzmann constant and $T$ is the temperature. For the levitated nanoparticle scenario (A), we note that this noise model can describe the regime where scattering off background gas molecules is the dominant noise source \cite{Epstein1924PR,Beresnev1990JFluidMech}. In the regime where recoil from dipole scattering of tweezer photons is the dominant mechanical noise source, the noise model would need some modification \cite{Seberson2020PRA}. However, we emphasize that the exact form of the oscillator's noise model is not important for the main results presented in this article.

Going to a description in terms of bright and dark modes, we find 
\begin{align}
\label{eq:xbEqs}
\dot{\hat{x}}_b & = \frac{\hat{p}_b}{m} \\
\dot{\hat{p}}_b & = -\gamma_b \hat{p}_b -\eta_m \, \hat{p}_d - m\omega_b^2 \hat{x}_b - m \bar{\omega}_m \delta_m  \hat{x}_d +  \hat{N}_b  \notag \\
& - \hbar \lambda_o \left(\hat{a} + \hat{a}^\dagger \right)  \\
\dot{\hat{x}}_d & = \frac{\hat{p}_d}{m} \\
\dot{\hat{p}}_d & = -\gamma_d \hat{p}_d -\eta_m \, \hat{p}_b - m\omega_d^2 \hat{x}_d - m \bar{\omega}_m \delta_m  \hat{x}_b  +  \hat{N}_d 
\label{eq:QLEMech4}
\end{align}
where we have defined bright and dark mode resonance frequencies and decay rates
\begin{align}
\label{eq:BrightDarkQuantities1}
\omega_b^2 & = \cos^2\phi \, \omega_x^2 + \sin^2\phi \, \omega_y^2 \\
\label{eq:BrightDarkQuantities2} \omega_d^2 & = \sin^2 \phi \, \omega_x^2 + \cos^2 \phi \, \omega_y^2 \\
\label{eq:BrightDarkQuantities3} \gamma_b & = \cos^2\phi \, \gamma_x + \sin^2\phi \, \gamma_y \\
\label{eq:BrightDarkQuantities4} \gamma_d & = \sin^2\phi \, \gamma_x + \cos^2\phi \, \gamma_y , 
\end{align}
the average mechanical resonance frequency
\begin{align}
\label{eq:AveMech}
\bar{\omega}_m & = \frac{1}{2} \left(\omega_x + \omega_y \right) ,
\end{align}
and where  
\begin{align}
\label{eq:CoupleDefs1}
\delta_m & =   \left( \omega_x - \omega_y \right) \sin 2\phi \\
\label{eq:CoupleDefs2} \eta_m & = \frac{1}{2} \left(\gamma_x - \gamma_y \right) \sin 2\phi
\end{align}
are parameters quantifying the coupling between the bright and dark modes. We have also defined the noise operators
\begin{align}
\label{eq:BrightNoise}
\hat{N}_b & = \cos \phi \, \hat{N}_x - \sin \phi \, \hat{N}_y \\
\label{eq:DarkNoise} \hat{N}_d & = \sin \phi \, \hat{N}_x + \cos \phi \, \hat{N}_y  .
\end{align}
Finally, we have the equation of motion for the cavity mode,
\begin{align}
\label{eq:QLECav}
\dot{\hat{a}} & = - \left(\frac{\kappa}{2} + i \Delta\right) \hat{a}  - i \lambda_o \hat{x}_b + \sqrt{\kappa} \,  \hat{\xi} .
\end{align}
Here, $\kappa$ denotes the bare energy decay rate due to coupling to the cavity mode's bath and we will assume $\kappa \gg \gamma_x, \gamma_y$ throughout this article. Assuming $\hbar \omega_c \gg k_B T$, the Gaussian operator $\hat{\xi}$ represents electromagnetic vacuum noise driving the cavity mode and satisfies
 \begin{align}
\label{eq:NoiseRel2a}
\langle \hat{\xi}[\omega] \hat{\xi}^\dagger[\omega'] \rangle  & =  2\pi \delta(\omega + \omega') \\
\label{eq:NoiseRel2b}
\langle \hat{\xi}^\dagger[\omega] \hat{\xi}[\omega'] \rangle  & = \langle \hat{\xi}[\omega] \hat{\xi}[\omega'] \rangle = 0 .
 \end{align}

We observe in Equation \eqref{eq:QLEMech4} that while the dark mechanical mode does not couple directly to the cavity mode, it does couple to the bright mechanical mode when $\delta_m \neq 0$. This is the case when both $\omega_y \neq \omega_x$, i.e., the harmonic trap is not rotationally symmetric, and $\phi \neq 0,\pi/2$, i.e., the cavity axis is not aligned with one of the principal axes of the trap. Additionally, we observe that the bright and dark modes can also be dissipatively coupled, quantified by the decay rate $\eta_m$, and that the two modes in general couple to correlated baths. In the special case $\gamma_x = \gamma_y$, we get $\eta_m = 0 $ and that $ \hat{N}_b$ and $\hat{N}_d$ are uncorrelated. We note that the latter would not be true if the noise model were modified to describe recoil from asymmetric dipole scattering. 

In Section \ref{sec:1DOM}, we will study the special case where $\delta_m = 0$ and where we can ignore the dissipative coupling $\eta_m$. In this case, the dark mode is {\it truly} dark, i.e., it does not couple to any of the other modes. The cavity mode and the bright mode then form a standard optomechanical system (with a 1D mechanical oscillator). In Section \ref{sec:2DOM}, we will consider the case where $\delta_m \neq 0$ and where the cavity mode effectively couples to the full 2D motion of the oscillator.

\section{Purity of a 1D oscillator coupled to an optical cavity mode}
\label{sec:1DOM}

In this Section, we consider $\delta_m = 0$, i.e., either that the harmonic trap is spherically symmetric ($\omega_x = \omega_y$) or that one of the principal axes of the trap is lined up with the optical cavity axis ($\sin 2\phi = 0$). With this assumption, there is no conservative interaction between the bright and dark mechanical modes. We also neglect the potential dissipative interaction between the mechanical modes which is exact for $\sin 2\phi = 0$ or $\gamma_x = \gamma_y$ in our model, but will generally also be a good approximation when the bright mode dissipation is dominated by its coupling to the cavity mode.

With these assumptions, we are now left with the standard optomechanical setup in which a single cavity mode couples to a single mechanical mode - the bright mode. We can solve the linear quantum Langevin equations in the frequency domain, which gives 
\begin{align}
\label{eq:QLEFour}
\hat{x}_b[\omega]  & = R_b(\omega) \hat{N}_{b,\mathrm{eff}}[\omega]
\end{align}
for the position operator of the bright mode, where we have defined a response function
\begin{align}
\label{eq:Response}
 & R_b(\omega) \\
 & = \frac{1}{ - i m \omega \gamma_b + m \left(\omega_b^2 - \omega^2 \right) - i \hbar \lambda_o^2 \left(\chi_c(\omega)  - \chi_c^\ast(-\omega) \right)}  \notag
\end{align}
and an effective noise operator
\begin{align}
\label{eq:EffNoise}
\hat{N}_{b,\mathrm{eff}}[\omega]  & =  \hat{N}_b[\omega]  - \sqrt{ \kappa} \hbar \lambda_o \left(\chi_c(\omega) \hat{\xi}[\omega] + \chi_c^\ast(-\omega) \hat{\xi}^\dagger[\omega] \right)
\end{align}
in terms of the bare cavity susceptibility
 \begin{equation}
\label{eq:CavSusc}
\chi_c(\omega) = \frac{1}{\kappa/2 - i(\omega - \Delta)} .
\end{equation}

In the following, we will be calculating the thermal occupation number $\bar{n}_b$ defined in the basis in which the reduced mechanical density matrix is thermal, given by \eqref{eq:BOccNum2}.  We will also compare this to the standard average phonon number $\bar{n}_{b,0}$, i.e., the average occupation number in the basis of eigenstates of the isolated oscillator Hamiltonian
 \begin{equation}
\label{eq:H0b}
\hat{H}_{b,0} = \frac{\hat{p}_b^2}{2m} + \frac{1}{2} m \omega_b^2 \hat{x}_b^2  .
\end{equation}
By defining the standard phonon annihilation operator
 \begin{equation}
\label{eq:PhononAnnihilationOperator}
\hat{b}_0  = \frac{1}{2} \left(\frac{\hat{x}_b}{x_{\mathrm{zpf},b,0}} + i \frac{\hat{p}_b}{p_{\mathrm{zpf},b,0}} \right) ,
\end{equation}
where
\begin{align}
\label{eq:xzpf0} x_{\mathrm{zpf},b,0} & = \sqrt{\frac{\hbar}{2m\omega_b}}  \quad , \quad
p_{\mathrm{zpf},b,0} = \sqrt{\frac{\hbar m\omega_b}{2}}
\end{align} 
are the magnitudes of position and momentum fluctuations in the ground state of the Hamiltonian \eqref{eq:H0b}, the average phonon number $\bar{n}_{b,0} = \langle \hat{b}_0^\dagger \hat{b}_0 \rangle$ can be calculated from \begin{align}
\label{eq:PhononNumberDef}
2 \bar{n}_{b,0}  + 1 & = \frac{1}{2} \left(\frac{\langle \hat{x}_b^2 \rangle}{x_{\mathrm{zpf},b,0}^2} +  \frac{\langle \hat{p}_b^2 \rangle}{p_{\mathrm{zpf},b,0}^2} \right) .
\end{align}

The annihilation operator $\hat{b}$ defined in Section \ref{sec:Purity} is related to the phonon operator $\hat{b}_0$ defined in \eqref{eq:PhononAnnihilationOperator} by
\begin{align}
\label{eq:AnnihilationRelation}
\hat{b} & =  \nu_+ \left(\hat{b}_0 - \bar{b}_0\right) + \nu_- \left(\hat{b}_0^\dagger - \bar{b}_0^\ast \right)
\end{align}
where 
\begin{align}
\label{eq:bBarDef}
\bar{b}_0 & =  \frac{1}{2} \left(\frac{\langle \hat{x}_b \rangle}{x_{\mathrm{zpf},b,0}} + i \frac{\langle \hat{p}_b \rangle}{p_{\mathrm{zpf},b,0}} \right) 
\end{align}
and the coefficients $\nu_\pm$ are defined as
\begin{align}
\label{eq:nuDef}
 \nu_\pm & = \frac{1}{1+e^{2i\theta}} \left(\frac{x_{\mathrm{zpf},b,0}}{x_{\mathrm{zpf},b}} \pm e^{i\theta}\frac{p_{\mathrm{zpf},b,0}}{p_{\mathrm{zpf},b}} \right) .
\end{align}
Throughout this Section, we will use that $\langle \hat{x}_b \rangle = \langle \hat{p}_b \rangle = 0$ in the model defined in Section \ref{sec:Model}, such that $\Delta \hat{x}_b = \hat{x}_b$ and $\Delta \hat{p}_b = \hat{p}_b$, in which case $\bar{b}_0 = 0$ and \eqref{eq:AnnihilationRelation} reduces to a {\it Bogoliubov transformation}. Finally, we note that our model also gives $\langle \{\hat{x}_b , \hat{p}_b \} \rangle = 0$, such that $\theta = 0$.

\subsection{Weak optomechanical coupling}
\label{sec:WC1D}

Let us first consider the weak coupling limit where the effective mechanical decay rate $\tilde{\gamma}_b$, to be defined below, is much smaller than the cavity linewidth $\kappa$. In this limit, the mechanical mode still behaves as a harmonic oscillator in thermal equilibrium, but with a renormalized resonance frequency $\tilde{\omega}_b$ and a renormalized linewidth $\tilde{\gamma}_b$. Assuming $\tilde{\gamma}_b \ll \tilde{\omega}_b$, we may approximate the response function by
\begin{align}
\label{eq:ResponseWeak}
 R_b(\omega) & = \frac{i \, \mathrm{sgn}(\omega)}{ 2 m \tilde{\omega}_b \left[\tilde{\gamma}_b/2 - i \, \mathrm{sgn}(\omega)\left(|\omega| - \tilde{\omega}_b \right) \right] }   
\end{align}
when implicitly defining the effective mechanical resonance frequency $\tilde{\omega}_b$ according to
\begin{align}
\label{eq:OptSpring}
\tilde{\omega}^2_b & = \omega_b^2 + \frac{\hbar \lambda_o^2}{m} \mathrm{Im} \left[\chi_c(\tilde{\omega}_b) - \chi_c^\ast(-\tilde{\omega}_b)\right] , 
\end{align}
showing that $\tilde{\omega}_b$ is shifted from its bare value $\omega_b$ due to the optical spring effect, and the effective linewidth
\begin{align}
\label{eq:gammaTilde}
\tilde{\gamma}_b & = \gamma_b + \frac{\hbar \lambda_o^2}{m \tilde{\omega}_b} \mathrm{Re} \left[\chi_c(\tilde{\omega}_b) - \chi_c^\ast(-\tilde{\omega}_b)\right] .
\end{align}

This limit leads to a simplified expression for the position spectral density 
\begin{align}
\label{eq:SxxDef}
S_{x_b x_b}(\omega) & = \int_{-\infty}^\infty \frac{d \omega'}{2\pi} \, \langle \hat{x}_b[\omega] \hat{x}_b[\omega'] \rangle ,
\end{align}
whose integral over all frequencies gives the position variance $\langle \hat{x}^2_b \rangle$. Since $S_{x_b x_b}(\omega)$ is appreciably non-zero only in narrow regions around $\omega = \pm \tilde{\omega}_b$, we may approximate $\omega \approx \pm \tilde{\omega}_b$ when using Equations \eqref{eq:NoiseRel} and \eqref{eq:CavSusc} to find a simplified expression for $S_{x_b x_b}(\omega)$. The spectral density then consists of two Lorentzians of width $\tilde{\gamma}_b$ centered at $\omega = \pm \tilde{\omega}_b$, whose frequency integral gives 
\begin{align}
\label{eq:Varx}
\langle \hat{x}_b^2 \rangle & = \frac{\hbar}{2 m \tilde{\omega}_b}\\
& \times \Bigg(2 \frac{\gamma_b n_B(\tilde{\omega}_b) + \kappa \hbar \lambda_o^2 |\chi_c(-\tilde{\omega}_b)|^2/(2 m \tilde{\omega}_b)  }{\tilde{\gamma}_b} + 1 \Bigg) \notag
\end{align}
with
\begin{align}
\label{eq:Planck}
 n_B(\omega) = \left(e^{\hbar \omega/(k_B T)} - 1 \right)^{-1} 
\end{align}
being the Planck distribution. The variance of the bright mode's momentum can also be calculated from the spectral density, giving
\begin{align}
\label{eq:Varp}
\langle \hat{p}_b^2 \rangle & = \int \frac{d\omega}{2\pi} m^2 \omega^2 S_{x_b x_b}(\omega) \approx m^2 \tilde{\omega}_b^2 \langle \hat{x}^2_b \rangle
\end{align}
when exploiting again that the spectral density $S_{x_b x_b}(\omega)$ is narrowly peaked at $\pm \tilde{\omega}_b$. 

From Equation \eqref{eq:xzpf}, 
this leads to the approximate relation
\begin{align}
\label{eq:zpfWeak}
 x_{\mathrm{zpf},b} = \sqrt{\frac{\hbar}{2m\tilde{\omega}_b}} 
\end{align}
and $p_{\mathrm{zpf},b} = \hbar/(2x_{\mathrm{zpf},b})$, which are indeed the position and momentum zero-point fluctuations for an isolated harmonic oscillator at resonance frequency $\tilde{\omega}_b$ in the ground state. Furthermore, Equation \eqref{eq:BOccNum2} gives 
\begin{align}
\label{eq:nmWeak}
\bar{n}_b = \frac{\gamma_b n_B(\tilde{\omega}_b) + \kappa \lambda_o^2 x_{\mathrm{zpf},b}^2 |\chi_c(-\tilde{\omega}_b)|^2 }{\tilde{\gamma}_b} 
\end{align}
for the bright mode's thermal occupation number. This matches the well-known result for the oscillator's average phonon number from the standard theory of optomechanics \cite{Wilson-Rae2007PRL,Marquardt2007PRL}, except that the zero point motion $x_{\mathrm{zpf},b}$, as well as the arguments of the bath occupation number \eqref{eq:Planck} and the cavity susceptibility \eqref{eq:CavSusc}, is defined in terms of the oscillator's {\it effective} resonance frequency. 

Equations \eqref{eq:zpfWeak} and \eqref{eq:nmWeak} are our first results. They show that the purity $\mu_b = 1/(2\bar{n}_b + 1)$ is directly related to the average phonon number in the weak coupling regime, but only when defined in terms of the oscillator's effective resonance frequency. For small optical spring shifts $|\tilde{\omega}_b - \omega_b| \ll \omega_b$, the error made by replacing $\tilde{\omega}_b \rightarrow \omega_b$ in \eqref{eq:nmWeak} is small. However, when the optical spring shift becomes comparable to or even larger than the bare frequency, e.g., as in more advanced experimental setups \cite{Corbitt2007PRL_2,Corbitt2007PRL}, the phonon number must be defined in terms of the effective frequency in order to reliably quantify state purity.

\subsection{Strong and ultrastrong optomechanical coupling}

Let us now consider the strong coupling limit in which the mechanical and cavity modes hybridize into resolvable normal modes \cite{Dobrindt2008PRL}, sometimes referred to as polariton modes \cite{Lemonde2013PRL,Ranfagni2021NatPhys} in analogy with solid-state systems featuring strong light-matter interaction. We start by writing the mechanical response function on the form
\begin{align}
R_b(\omega) = \frac{\chi^{-1 }_c(\omega) \chi^{-1 \, \ast}_c(-\omega)}{m (\omega - z_-)(\omega + z_-^\ast)(\omega - z_+) (\omega + z_+^\ast)} ,
\end{align}
and we express the poles 
\begin{align}
z_\pm \equiv \omega_\pm - i \frac{\kappa_\pm}{2}
\end{align} 
in terms of normal mode frequencies $\omega_\pm$ and linewidths $\kappa_\pm$. 

For simplicity, we assume $\Delta = \omega_b$ in this subsection. We also assume that the normal modes are well separated in frequency, meaning  $\kappa_\pm  \ll \omega_+ - \omega_- $, 
and that the lower frequency normal mode is a high-$Q$ oscillator, i.e, $\kappa_- \ll \omega_-$. 
One can then write down approximate expressions for the normal mode resonance frequencies
\begin{align}
\label{eq:PolFreqs1D}
\omega_\pm = \omega_b \sqrt{1 \pm \frac{2G_o}{\omega_b}}
\end{align} 
where
\begin{align}
\label{eq:GoDef}
G_o = \lambda_o x_{\mathrm{zpf},b,0}  
\end{align} 
is the standard definition of the enhanced optomechanical coupling rate \cite{Aspelmeyer2014RMP}. To reach \eqref{eq:PolFreqs1D}, we have neglected relative corrections of order $(\kappa/G_o)^2 , (\kappa/\omega_b)^2$
in line with the strong coupling assumption. For the normal mode linewidths, we get
\begin{align}
\label{eq:kappapm}
\kappa_\pm = \frac{\kappa}{2}
\end{align}  
in the limit $\gamma_b/\kappa \rightarrow 0$. 

With the above assumptions, the solution \eqref{eq:QLEFour} gives approximately
\begin{align}
\label{eq:xFluctStrongCoup}
\langle \hat{x}_b^2 \rangle =  x^2_{\mathrm{zpf},-} \left(2 \bar{n}_- +1\right) + x^2_{\mathrm{zpf},+}  \left(2 \bar{n}_+ + 1\right) 
\end{align} 
when defining \cite{Lemonde2013PRL}
\begin{align}
\label{eq:npmStrong}
\bar{n}_\pm = \frac{\gamma_b n_B(\omega_\pm)/2 + \kappa (\omega_\pm - \omega_b)^2/(8 \omega_b \omega_\pm)}{\kappa_\pm} 
\end{align} 
and
\begin{align}
\label{eq:xzpfpmDef}
x_{\mathrm{zpf},\pm} = \sqrt{\frac{\hbar}{4m \omega_\pm}} .
\end{align}
The result \eqref{eq:xFluctStrongCoup} can be interpreted as the mechanical fluctuations originating from the two normal modes, with average polariton occupation numbers $\bar{n}_\pm$ and where $x_{\mathrm{zpf},\pm}$ are the mechanical fluctuations when $n_\pm = 0$.
\footnote{One should generally distinguish the polariton occupation numbers $\bar{n}_\pm$, defined here in the context of a 1D mechanical oscillator, from the thermal occupation numbers $\bar{m}, \bar{n}$ defined in Section \ref{sec:2DGauss} for a 2D oscillator. However, if one were to consider the combined bright and cavity modes as a 2D oscillator, $\bar{n}_\pm$ would correspond to $\bar{m},\bar{n}$ in the strong coupling regime.}. Similarly, the variance of the oscillator momentum becomes 
\begin{align}
\langle \hat{p}_b^2 \rangle = p^2_{\mathrm{zpf},-} \left(2 \bar{n}_- +1\right) + p^2_{\mathrm{zpf},+}  \left(2 \bar{n}_+ + 1\right) 
\end{align} 
with
\begin{align}
\label{eq:pzpfpmDef}
p_{\mathrm{zpf},\pm} = \sqrt{\frac{\hbar m \omega_\pm}{4}} .
\end{align} 

The thermal occupation number for the bright mode defined by Equation \eqref{eq:BOccNum2} can now be written 
\begin{align}
\label{eq:ThermOccStrongCoupling}
2 \bar{n}_b  + 1 & = \frac{1}{2} \Big[(2 \bar{n}_- +1)^2  + (2 \bar{n}_+ +1)^2 \\
& + \frac{2\omega_b^2 }{\omega_+ \omega_-}(2 \bar{n}_- +1)(2 \bar{n}_+ +1) \Big]^{1/2}  . \notag 
\end{align} 
We note that for strong but not ultrastrong coupling, meaning $\kappa \ll G_o \ll \omega_b$, we have the approximation $\bar{n}_b = (\bar{n}_- + \bar{n}_+)/2$, i.e., the average of the normal mode occupation numbers. 

In the strong coupling regime we have considered in this subsection, the bright mode position spectral density, which is accessible by heterodyne photodetection of the cavity output field, consists of four well-separated Lorentzian peaks at frequencies $\pm \omega_\pm$. The average normal mode occupation numbers $\bar{n}_\pm$ can thereby be accessed from the asymmetries of the peak heights at positive and negative frequencies, similarly to how mechanical occupation numbers can be determined in weakly coupled optomechanical systems \cite{Weinstein2014PRX,Purdy2015PRA,Underwood2015PRA,Sudhir2017PRX,Tebbenjohanns2020PRL}. This method of finding $n_\pm$ from the peak height ratios means that the thermal occupation number $\bar{n}_b$ in Equation \eqref{eq:ThermOccStrongCoupling} can be determined without having to integrate the spectral density over all frequencies. This also alleviates the need for calibrating the detector signal to oscillator position, since the sideband peak height ratios are gain independent as long as the gain is the same at all sideband frequencies.

It is worth comparing the thermal occupation number in \eqref{eq:ThermOccStrongCoupling} to the average occupation number $\bar{n}_{b,0}$ defined in Equation \eqref{eq:PhononNumberDef}. In the strong coupling regime discussed above and for $\Delta = \omega_b$, we find
\begin{align}
\label{eq:PhononNumberSC}
2 \bar{n}_{b,0}  + 1 & = \frac{1}{4} \sum_{\sigma = \pm} \frac{\omega_b^2 + \omega_\sigma^2}{\omega_b \omega_\sigma} \left(2 \bar{n}_\sigma + 1\right) .
\end{align}
For $G_o \ll \omega_b$, such that $|\omega_b - \omega_\pm| \ll \omega_b$, $\bar{n}_b$ and $\bar{n}_{b,0}$ are approximately equal. However, as is clear from Equations \eqref{eq:ThermOccStrongCoupling} and \eqref{eq:PhononNumberSC}, and will be shown graphically below, they differ significantly in the ultrastrong coupling regime $G_o \sim \omega_b$ where the concepts of purity and average phonon number are not closely related.

\subsection{Exact results in the quantum backaction limit}

After having explored the regimes of weak and strong coupling, we now wish to derive an expression for the average thermal occupation number $\bar{n}_b$, and thus the quantum state purity $\mu_b$, for arbitrary optomechanical coupling rate $G_o$ and detuning $\Delta > 0$. To avoid unwieldy expressions, however, we will consider the limit where the motion of the oscillator is dominated by electromagnetic vacuum noise due to its coupling to the cavity mode (as opposed to the noise from its own bath), i.e., the quantum backaction limit \cite{Wilson-Rae2007PRL,Marquardt2007PRL}. In other words, we will calculate the {\it minimal} $\bar{n}_b$ achievable by optomechanical cooling for a general coupling rate $G_o$. This is relevant in the regime where
\begin{align}
\label{eq:QBLimit}
\frac{\gamma_b k_B T}{\hbar \omega_b} \ll  \mathrm{min}\left(\frac{G_o^2}{\kappa} \, , \, \kappa \right),
\end{align}
which is a well-known requirement for ground state cooling of the mechanical mode \cite{Wilson-Rae2007PRL,Marquardt2007PRL}. 

From the presence of $n_B(\omega_-)$ in the normal mode occupancy $\bar{n}_-$ (see Equation \eqref{eq:npmStrong}), one might worry that in the strong coupling regime, the temperature $T$ should be compared to the lower normal mode frequency $\omega_-$, not $\omega_b$ as in \eqref{eq:QBLimit}. However, one should note that the second term in the numerator of \eqref{eq:npmStrong}, originating from the electromagnetic vacuum noise $\hat{\xi}$, also scales inversely with $\omega_-$. Thus, even in the ultrastrong coupling regime where $\omega_- \ll \omega_b$, the motion is dominated by the electromagnetic vacuum noise (i.e., radiation pressure shot noise) as long as \eqref{eq:QBLimit} is satisfied.

Starting from the equations of motion in the time domain (see Section \ref{sec:Model}) and setting $\gamma_b = 0$ to explore the quantum backaction limit, we derive the steady state expectation values
\begin{align}
\label{eq:xSqExact1D}
\langle \hat{x}_b^2 \rangle & = \frac{\hbar}{4m\Delta} \left(1 + \frac{(\kappa/2)^2 + \Delta^2}{\omega_b^2 - 2g_o^2} \right), \\
\label{eq:pSqExact1D}
\langle \hat{p}_b^2 \rangle & = \frac{\hbar m \left[ (\kappa/2)^2 + \Delta^2 + \omega_b^2 \right]}{4 \Delta} ,
\end{align}
and $\langle \{\hat{x}_b , \hat{p}_b \} \rangle = 0$, when defining
\begin{align}
\label{eq:gDef}
g_o^2 = \frac{\hbar \lambda^2_o \Delta}{m \left[(\kappa/2)^2 + \Delta^2\right]} = \frac{2 G_o^2 \Delta \omega_b}{ (\kappa/2)^2 + \Delta^2} .
\end{align}
These expectation values can then be used to calculate the thermal occupation number $\bar{n}_b$ according to \eqref{eq:BOccNum2}, giving
\begin{align}
\label{eq:nmExact}
2 \bar{n}_b & + 1 \\
& = \sqrt{\frac{\left[(\kappa/2)^2 + \Delta^2 + \omega_b^2 - 2g_o^2\right]\left[ (\kappa/2)^2 + \Delta^2 + \omega_b^2 \right]}{4\Delta^2 \left(\omega_b^2 - 2g_o^2\right)}} , \notag
\end{align}
 and thereby the purity \eqref{eq:PurityThermal} in the quantum backaction limit. It is straightforward to verify that in the weak-coupling regime $g_o \ll \omega_b$, this reproduces the well-known result  
\begin{align}
\label{eq:nmMin}
\lim_{G_o \rightarrow 0} \bar{n}_{b}  = \frac{(\kappa/2)^2 + (\Delta - \omega_b)^2}{4 \omega_b \Delta} ,
 \end{align}
for the minimal average phonon number \cite{Wilson-Rae2007PRL,Marquardt2007PRL}. It is clear from \eqref{eq:nmExact} that this is the minimal $\bar{n}_b$ achievable in this setup. We also note that expanding Equation \eqref{eq:nmExact} to second order in $g_o$ gives agreement with previously reported results \cite{Dobrindt2008PRL} for the minimal average phonon number valid in the strong, but not ultrastrong, coupling limit.

\begin{figure}[htb]
\includegraphics[width=.99\columnwidth]{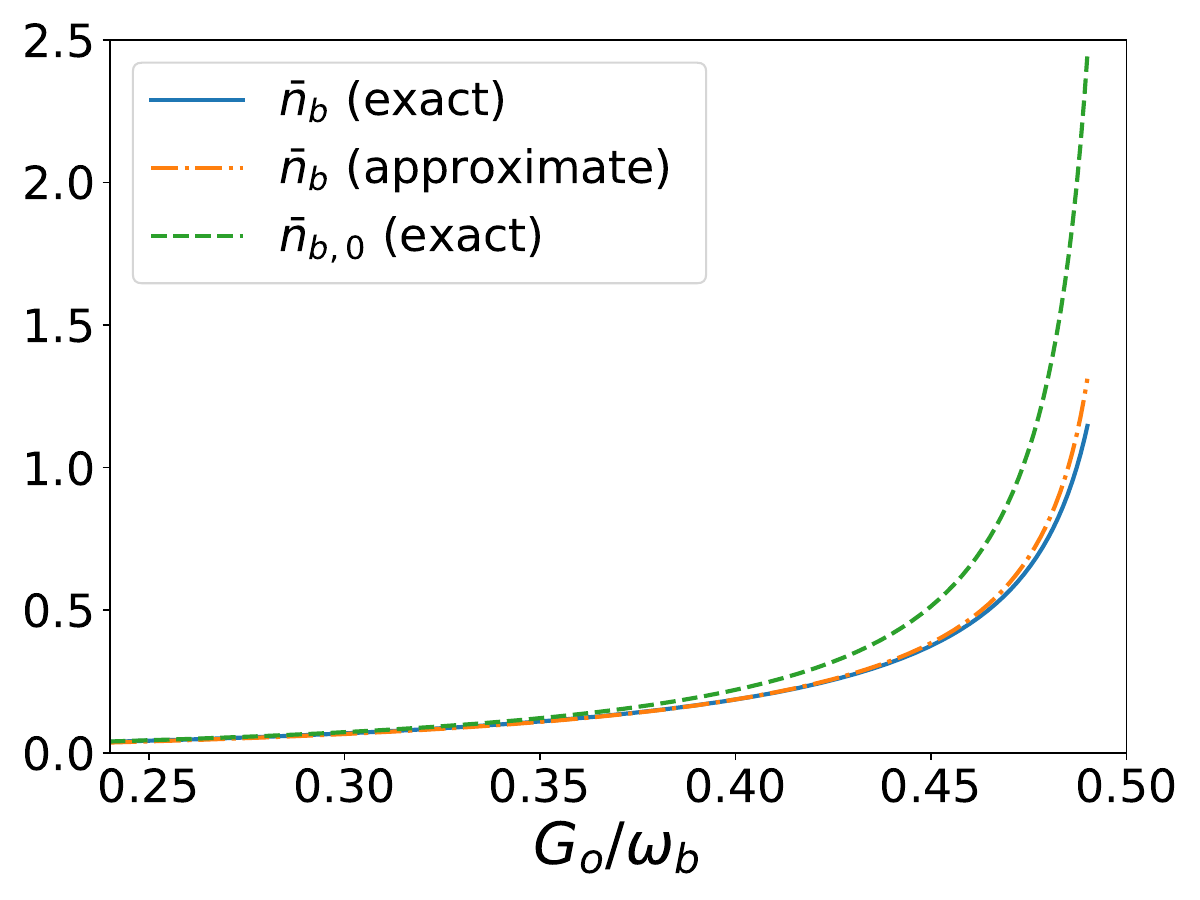}
\caption{Thermal occupation number $\bar{n}_b$ from \eqref{eq:nmExact} (solid, blue line) and average phonon number $\bar{n}_{b,0}$ from \eqref{eq:PhononNumberDef} (dashed, green line) as function of coupling rate $G_o$ in the quantum backaction limit where we have set $\gamma_b = 0$. We have chosen $\Delta = \omega_b$ and $\kappa/\omega_b = 0.2$. We also plot the approximate strong coupling result for $\bar{n}_b$ from \eqref{eq:ThermOccStrongCoupling} (dash-dotted, orange line).}
\label{fig:Plotnm1D}
\end{figure}

In Figure \ref{fig:Plotnm1D}, we plot $\bar{n}_b$ as given by the exact expression \eqref{eq:nmExact} in the strong coupling regime $G_o > \kappa$ when choosing $\Delta = \omega_b$ and $\kappa/\omega_b = 0.2$. We also plot the approximate result for $\bar{n}_b$ given by \eqref{eq:ThermOccStrongCoupling}, which is valid in the strong coupling regime and for lower normal mode frequency $\omega_- \gg \kappa$. Finally, we plot the average phonon number $\bar{n}_{b,0}$ from inserting Equations \eqref{eq:xSqExact1D} and \eqref{eq:pSqExact1D} into the definition \eqref{eq:PhononNumberDef}. While $\bar{n}_b$ and $\bar{n}_{b,0}$ coincide for small values of coupling strength $G_o$, they differ in the ultrastrong coupling regime where $G_o$ is comparable to the bare resonance frequency $\omega_b$. The thermal occupation number $\bar{n}_{b}$ is significantly smaller than the average phonon number $\bar{n}_{b,0}$, which tells us that the average phonon number is not a good indicator of the purity of the quantum state of the mechanical mode, i.e., that the state is non-thermal in the phonon basis. We also observe that the approximate strong coupling result for $\bar{n}_b$ matches the exact result well, except for values of $G_o$ close to $\omega_b/2$ where the assumption $\omega_- \gg \kappa$ breaks down since $\omega_- \rightarrow 0$.

We can also calculate the parameter $M\Omega$, defined by Equation \eqref{eq:OmegaComplexTildelambdaDef}, which enters the position space eigenstates \eqref{eq:Psin} in the probabilistic mixture representation of the mechanical mode's Gaussian state, giving
\begin{align}
\label{eq:xzpfExact}
M \Omega =m \sqrt{ \frac{ \left[(\kappa/2)^2 + \Delta^2 + \omega_b^2 \right] \left(\omega_b^2 - 2g_o^2\right)}{\left[(\kappa/2)^2 + \Delta^2 + \omega_b^2 - 2g_o^2\right]} } .
\end{align}
This reduces to $m \omega_b$ for $G_o = 0$, as expected. In the weak coupling limit $G_o \ll \kappa$, we find that $M\Omega \approx m \tilde{\omega}_b$ in the limit when the frequency shift $|\tilde{\omega}_b - \omega_b|$ due to the optical spring far exceeds the mechanical linewidth $  \tilde{\gamma}_b$, which is the case when $\kappa \gg \omega_b$. However, note that for general $\kappa/\omega_b$, there are additional, small corrections to $M\Omega$ of order $m \tilde{\gamma}_b$ which was neglected in Equations \eqref{eq:Varp} and \eqref{eq:zpfWeak}.

\section{Purity of a 2D oscillator coupled to an optical cavity mode}
\label{sec:2DOM}

We now move on to consider the model defined in Section \ref{sec:Model} for $\delta_m \neq 0$, i.e., the situation where the cavity mode effectively couples to the full, two-dimensional motion of the oscillator. For convenience, we introduce the rate
\begin{align}
\label{eq:GmDef}
G_m =   \frac{\bar{\omega}_m \delta_m}{2 \sqrt{\omega_b \omega_d}} 
\end{align}
which is a measure of the coupling between the bright and dark mechanical modes. We note that the model is defined such that $\langle \hat{x}_b \rangle = \langle \hat{p}_b \rangle = \langle \hat{x}_d \rangle = \langle \hat{p}_d \rangle = 0$. 

To determine the purity of the quantum state of the two-dimensional oscillator from Equation \eqref{eq:Purity2D}, one would need to measure not only the bright mode's position fluctuations $\hat{x}_b$, but also the dark mode's fluctuations $\hat{x}_d$. While the bright mode is directly accessible through photodetection of the cavity output field, the dark mode is (per definition) not. However, the fluctuations of the dark mode are in principle accessible through detection of light scattered orthogonally to the cavity axis.

\subsection{Rotating wave approximation}
\label{sec:RWA}

To gain insight, we will start by applying the rotating wave approximation, which means that we exclude ultrastrong coupling and consider the resolved sideband limit $\kappa/\omega_b \rightarrow 0$. To simplify, we specialize to isotropic friction, i.e., $\gamma_x = \gamma_y$, and thus set $\eta_m = 0$. In terms of the standard phonon annihilation operator $\hat{b}_0$ for the bright mode, defined in Equation \eqref{eq:PhononAnnihilationOperator}, and the phonon annihilation operator $\hat{d}_0$ for the dark mode, defined similarly, the equations of motion then become
\begin{align}
\label{eq:RWA}
\dot{\hat{a}} & = - \left(\frac{\kappa}{2} + i \Delta \right) \hat{a} - i G_o \hat{b}_0 + \sqrt{\kappa} \, \hat{\xi}  \\
\dot{\hat{b}}_0 & = - \left(\frac{\gamma_b}{2} + i \omega_b \right) \hat{b}_0 - i G_o \hat{a} - i G_m  \hat{d}_0 + \sqrt{\gamma_b} \, \hat{\zeta}_b \\
\dot{\hat{d}}_0 & = - \left(\frac{\gamma_d}{2} + i \omega_d \right) \hat{d}_0 - i G_m  \hat{b}_0 + \sqrt{\gamma_d} \, \hat{\zeta}_d ,
\end{align}
where
\begin{align}
\label{eq:etaDef}
\langle \hat{\zeta}_i^\dagger(t) \hat{\zeta}_j(t') \rangle & =  n_B(\omega_i) \delta_{ij} \delta(t-t') \\
\langle \hat{\zeta}_i(t) \hat{\zeta}^\dagger_j(t') \rangle & =  \left[n_B(\omega_i) + 1\right] \delta_{ij} \delta(t-t') .
\end{align}
To simplify further, we will also consider a rotation angle of $\phi = \pi/4$ for the harmonic trap (see Figure \ref{fig:Setup}), giving $\omega_d = \omega_b$, and a laser detuning $\Delta = \omega_b$, such that all three modes are resonant. The rotating wave approximation is then tantamount to the assumptions $\kappa, G_o, G_m \ll \omega_b$. Note that $\phi = \pi/4$ also gives $\gamma_b = \gamma_d$, and that we define $\gamma_\mathrm{tot} = \gamma_b + \gamma_d = 2 \gamma_b$.

Finally, we consider the regime of large optomechanical cooperativity
\begin{align}
\label{eq:CoptDef}
C_o = \frac{4 G_o^2}{\kappa \gamma_\mathrm{tot}} \gg 1 ,
\end{align}
which is the regime we primarily are interested in, and we assume $G_o^2 \gg G^2_m \gamma_\mathrm{tot}/\kappa $.

\begin{figure}[htb]
\includegraphics[width=.99\columnwidth]{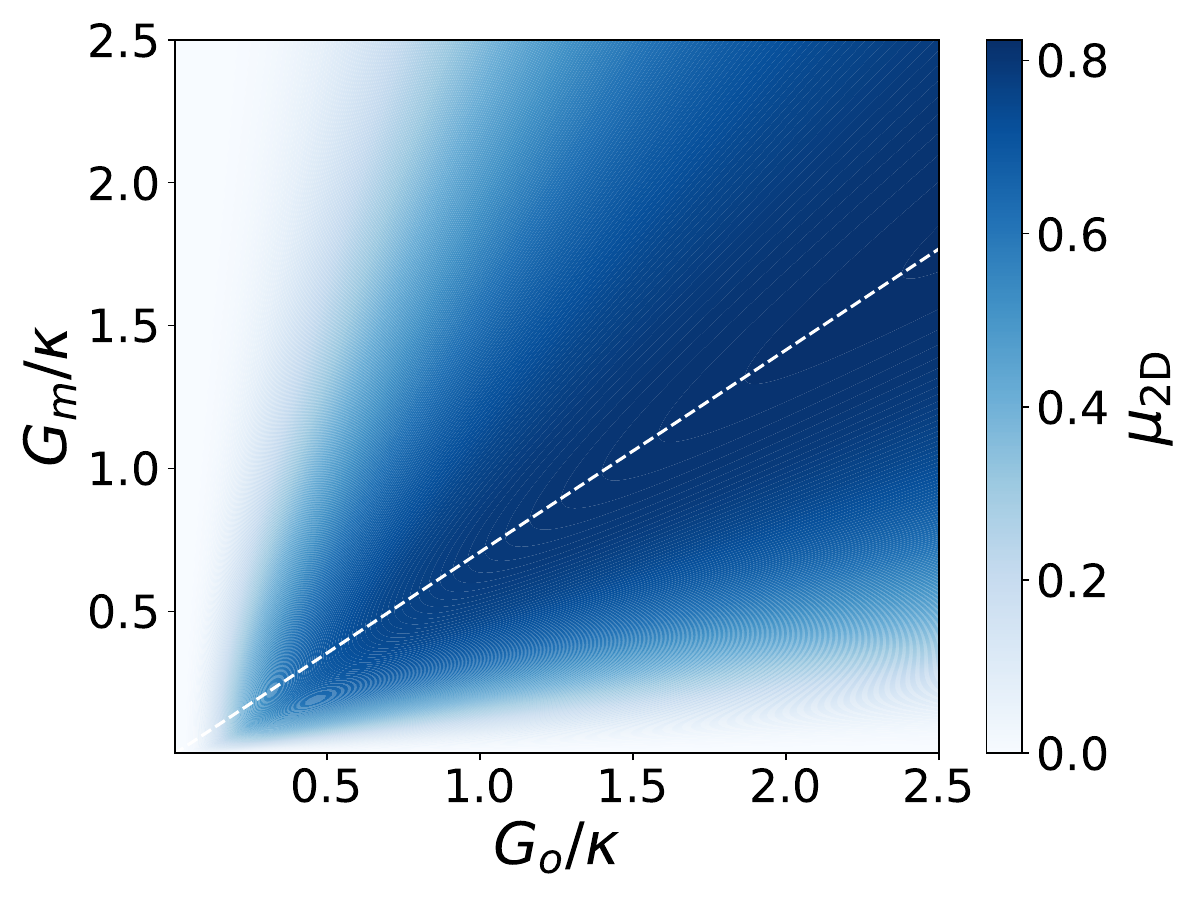}
\caption{Density plot of the state purity $\mu_\mathrm{2D}$ for the two-dimensional oscillator as a function of the optomechanical coupling rate $G_o$ and the mechanical coupling rate $G_m$ in the special cases $\gamma_x = \gamma_y$ and $\phi = \pi/4$, and with the assumptions $\kappa, G_o, G_m \ll \omega_b$. We have used the parameters $\gamma_\mathrm{tot}/\kappa = 1.0 \cdot 10^{-9}$, and $\gamma_\mathrm{tot} n_B(\omega_b)/\kappa = 0.05$. The dashed white line is given by $G_o/G_m = \sqrt{2}$, which is the coupling ratio that maximizes purity given our assumptions.}
\label{fig:PlotPurity2DRWA}
\end{figure}

With the above simplifications, the inverse purity of the mechanical oscillator's quantum state becomes
\begin{align}
\label{eq:Purity2DRWA}
\mu_\mathrm{2D}^{-1} =  \left(2 \bar{n}_{b,0} + 1\right) \left(2 \bar{n}_{d,0} + 1\right)  - 4|\langle \hat{b}_0^\dagger \hat{d}_0 \rangle |^2
\end{align}
according to Equation \eqref{eq:Purity2D}, where $\bar{n}_{b,0} = \langle \hat{b}_0^\dagger \hat{b}_0 \rangle$, $\bar{n}_{d,0} = \langle \hat{d}_0^\dagger \hat{d}_0 \rangle$, and we have used that $\mathrm{Re} \langle \hat{b}^\dagger_0 \hat{d}_0 \rangle = 0$ when $\Delta = \omega_b = \omega_d$. In Figure \ref{fig:PlotPurity2DRWA}, we plot the purity versus the two coupling rates $G_o, G_m$ found from solving the equations of motion above with the stated assumptions, where we have used parameters inspired by the levitated nanoparticle setup \cite{Ranfagni2022PRR,Piotrowski2022}.

In order for the dark mode to be cooled to a degree comparable to the bright mode, it is clear that the rate of coupling between the mechanical modes $G_m$ must be comparable to the optomechanical coupling rate $G_o$. Assuming also $ G_m^2 \gg G_o^2 \gamma_d/\kappa$ and $\bar{n}_{b,0}, \bar{n}_{d,0}  \ll 1$, we find that the purity is maximized when 
\begin{align}
\label{eq:EqOcc}
 G_m^2 = \frac{G_o^2}{2} 
\end{align}
with our assumptions, in which case its inverse can be approximated by
\begin{align}
\label{eq:Purity2DRWA_Max}
\mu_\mathrm{2D}^{-1}  & \approx 1 + 2 \left(\bar{n}_{b,0} + \bar{n}_{d,0} \right)\\
& =  1 + 4 n_B(\omega_b) \left( \frac{1}{C_o} +  \frac{\gamma_\mathrm{tot}}{\kappa} \right)  . \notag
\end{align}
We observe that a high purity, i.e., $\mu_\mathrm{2D}$ close to unity, requires $\gamma_\mathrm{tot} n_B(\omega_d) \ll \kappa$, as usual for optomechanical ground state cooling \cite{Wilson-Rae2007PRL,Marquardt2007PRL}. In the weak coupling regime $G_o \ll \kappa$, it additionally requires a cooperativity $C_o \gg n_B(\omega_b)$.

\subsection{Exact results in the quantum backaction limit}

While the rotating wave approximation applied in subsection \ref{sec:RWA} works well both in the weak and strong coupling regime, it fails in the ultrastrong coupling regime where the coupling rates $G_o$ and $G_m$ are comparable to the frequencies $\omega_b, \omega_d, \Delta$. While solving the general problem exactly gives unwieldy expressions, we again find an exact expression for the state purity in the quantum backaction limit, i.e., when assuming that the 2D mechanical motion is dominated by its coupling to the cavity mode and thus ignoring the mechanical baths.

Solving the equations of motion in Section \ref{sec:Model} when setting $\gamma_x = \gamma_y = 0$, such that $\gamma_b = \gamma_d = \eta_m = 0$, gives $\langle \{\hat{x}_b , \hat{p}_b \} \rangle = \langle \{\hat{x}_d , \hat{p}_d \} \rangle = \langle \hat{x}_b \hat{p}_d \rangle =  \langle \hat{x}_d \hat{p}_b \rangle = 0$ and
\begin{align}
\label{eq:2D_QBAL1}
\langle \hat{x}_b^2 \rangle & = \frac{\hbar}{4 m \Delta} \left( 1 + \frac{\left[(\kappa/2)^2  + \Delta^2 \right] \omega_d^2}{\left(\omega_b^2 - 2g_o^2 \right)\omega_d^2 -  \bar{\omega}_m^2 \delta_m^2} \right) 
\end{align}
\begin{align}
\label{eq:2D_QBAL2}
\langle \hat{x}_d^2 \rangle & = \frac{\hbar}{4 m \Delta} \left( 1 + \frac{\left[(\kappa/2)^2  + \Delta^2 \right] \left(\omega_b^2 - 2g_o^2\right)}{\left(\omega_b^2 - 2g_o^2 \right)\omega_d^2 -  \bar{\omega}_m^2 \delta_m^2} \right)  
\end{align} 
\begin{align}
\label{eq:2D_QBAL3}
\langle \hat{p}_i^2 \rangle & =  \frac{\hbar m  \left[(\kappa/2)^2  + \Delta^2  + \omega_i^2\right]}{4\Delta}   \ , \ i = b,d 
\end{align} 
\begin{align}
\label{eq:2D_QBAL4}
\langle \hat{x}_b \hat{x}_d \rangle & = - \left(\frac{\hbar}{4 m \Delta} \right) \frac{\left[(\kappa/2)^2  + \Delta^2 \right] \bar{\omega}_m \delta_m}{\left(\omega_b^2 - 2g_o^2 \right)\omega_d^2 -  \bar{\omega}_m^2 \delta_m^2} 
\end{align} 
\begin{align}
\label{eq:2D_QBAL5}
\langle \hat{p}_b \hat{p}_d \rangle & = \frac{\hbar m \bar{\omega}_m \delta_m}{4 \Delta} .
\end{align} 
These expressions provide an analytical result for the quantum backaction limit of the purity when inserted into Equation \eqref{eq:Purity2D}.

\begin{figure}[htb]
\includegraphics[width=.99\columnwidth]{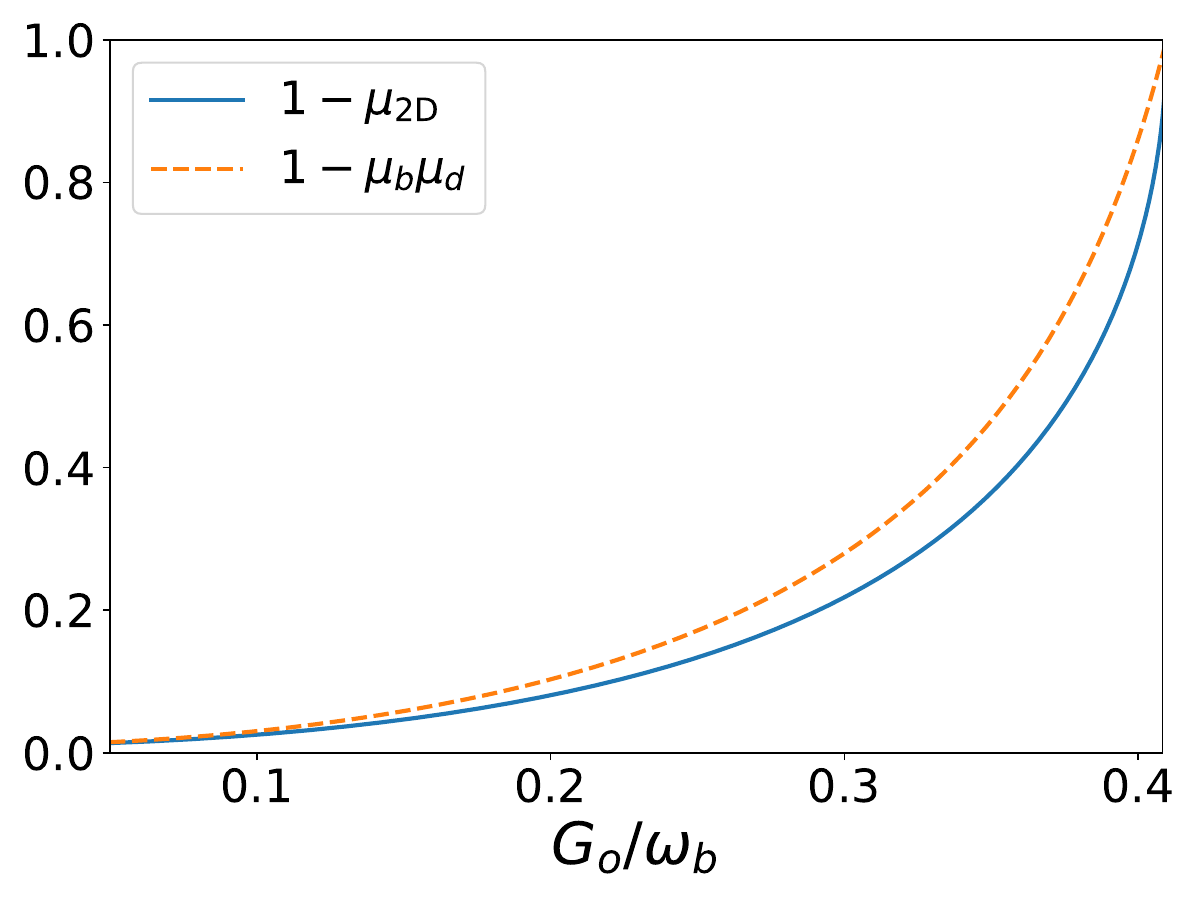}
\caption{Pure state deviation $1 - \mu_\mathrm{2D}$ of the two-dimensional mechanical oscillator, and $1 - \mu_b \mu_d$ given by the purities of the reduced states of the one-dimensional bright and dark modes, versus optomechanical coupling strength. We have used $\Delta = \omega_d = \omega_b$, $\kappa/\omega_b = 0.2$, $G_o/G_m = \sqrt{2}$, and assumed to be in the quantum backaction limit.} 
\label{fig:Purity2D_QBAL}
\end{figure}
One should note that the exact purity $\mu_\mathrm{2D}$ for the two-dimensional oscillator's state in general differs from the product of the separate state purities of the 1D bright and dark modes given by
\begin{align}
\label{eq:Purity2D1D1D}
 \mu_b \mu_d = \frac{\hbar^2}{4 \sqrt{\langle \hat{x}_b^2 \rangle \langle \hat{p}_b^2 \rangle  \langle \hat{x}_d^2 \rangle \langle\hat{p}_d^2 \rangle }} .
\end{align} 
In Figure \ref{fig:Purity2D_QBAL}, we compare the purity $\mu_\mathrm{2D}$ of the 2D mechanical state to the product $\mu_b \mu_d $. We observe that they differ in the ultrastrong coupling regime, indicating that the bright and dark modes are correlated. This shows that, in general, the complete characterization of a 2D oscillator cannot be limited to the occupation numbers along two orthogonal axes. \\

\section{Conclusion}
\label{sec:Conclusion}

In this article, we have argued that the thermal occupation number $\bar{n}$ of a one-dimensional oscillator in a Gaussian state should be defined with respect to the Fock basis in which the oscillator's quantum state is thermal. This has several advantages compared to the standard average phonon number. Firstly, it can be calculated from observable expectation values of position and momentum fluctuations without having to refer to the oscillator's confining potential. Secondly, it is directly related to state purity for all Gaussian states. In particular, for states that are squeezed thermal states in the phonon basis, it equals the number of Bogoliubons that measures the deviation from the {\it squeezed} vacuum \cite{Kronwald2013PRA_2}. Finally, it is also a meaningful quantity for non-Gaussian states, since, if viewing Equation \eqref{eq:BOccNum2} as a definition, $\bar{n}$ is a measure of deviation from a minimal uncertainty state. 

We studied optomechanical sideband cooling of a one-dimensional mechanical oscillator and showed that the deviation between the thermal occupation number we defined and the standard phonon number is most pronounced in the regime of ultrastrong coupling, i.e., where the optomechanical coupling rate is comparable to the bare mechanical resonance frequency, or when the oscillator's resonance frequency is strongly renormalized. However, since average phonon numbers in experiments are sometimes reported to the percent level accuracy, even small deviations between the two occupation numbers are noteworthy. We also derived an exact analytical expression for the minimal thermal occupation number achievable with sideband cooling for arbitrary coupling strength.

We have also argued that for higher-dimensional oscillators in Gaussian states, the use of quantum state purity to quantify the quantum character of the state is preferable to average phonon numbers along arbitrary directions. For Gaussian states, the purity is inversely proportional to the square root of the determinant of the covariance matrix and thus directly accessible through measurements of quadratic expectation values of positions and momenta. We considered sideband cooling in cavity optomechanics with two-dimensional mechanical oscillators, relevant to experiments with levitated nanoparticles, and derived an exact analytical expression for the maximal purity achievable in this setup. Finally, we note that an alternative quantifier of mixedness could be von Neumann entropy, which may provide a richer characterization scheme in the two-dimensional case as it then also depends on a second invariant of the covariance matrix in addition to the determinant \cite{Serafini2004JPhysB}.

\begin{acknowledgments}
We thank Francesco Massel for helpful comments.
\end{acknowledgments}

%


\newpage

\appendix*
\section{Overview of symbols}
\label{app:Symbols}
\onecolumngrid

\begin{table}[h]
\caption{\label{tab:Symbols} Overview of symbols used in the article.}
\begin{ruledtabular}
    \begin{tabular}{ c p{12cm} c}
    
    {\bf Symbol} & {\bf Description} & {\bf Relevant equations} \\ \hline
    $\hat{x}, \hat{y}$ & Components of position operator along the principal axes of the harmonic trap &   \\ 
    $\hat{p}_x , \hat{p}_y$ & Canonical conjugate momentum operators of $\hat{x}$, $\hat{y}$ &  \\ 
    $\hat{x}_b$, $\hat{x}_d$ & Bright/dark mode position operator & \eqref{eq:BrightDef} , \eqref{eq:DarkDef} \\
    $\hat{p}_b$, $\hat{p}_d$ & Canonical conjugate momentum operators of $\hat{x}_b$, $\hat{x}_d$ &    \\
    $\hat{a}$ & Photon annihilation operator & \\
    $\Delta \hat{o}$ & Fluctuation of operator $\hat{o}$ & \eqref{eq:FluctDef} \\
    $ m $ & Mass of mechanical oscillator & \\
    $ \phi $ & Angle between principal axis and cavity axis &  \\
    $ \omega_x , \omega_y $ & Eigenfrequencies along the principal axes of the harmonic trap &  \\
    $ \omega_b , \omega_d $ & Eigenfrequencies of the uncoupled bright and dark modes & \eqref{eq:BrightDarkQuantities1} , \eqref{eq:BrightDarkQuantities2} \\
    $ \gamma_b , \gamma_d $ & Bare energy decay rate for the bright and dark modes & \eqref{eq:BrightDarkQuantities3} , \eqref{eq:BrightDarkQuantities4} \\
    $ \kappa $ &  Bare energy decay rate for the cavity mode & \\
    $ T  $ & Temperature & \\
    $  k_B $ & Boltzmann's constant &  \\
    $  \hbar $ & Planck's reduced constant & \\
    $ \hat{N}_b , \hat{N}_d$ & Intrinsic noise operators for bright and dark modes & \eqref{eq:BrightNoise} ,  \eqref{eq:DarkNoise} , \eqref{eq:NoiseRel} \\
    $\hat{\xi} $ & Electromagnetic vacuum noise driving the cavity & \eqref{eq:NoiseRel2a} , \eqref{eq:NoiseRel2b} \\  
    $\lambda_o$ & Optomechanical coupling constant (with dimensions frequency/length) &  \\
    $ \bar{\omega}_m $ & Average of mechanical frequencies $\omega_x$ and $\omega_y$ & \eqref{eq:AveMech}  \\
    $ \delta_m $ & Measure of conservative coupling between bright and dark modes & \eqref{eq:CoupleDefs1}\\
    $ \eta_m $ & Measure of dissipative coupling between bright and dark modes & \eqref{eq:CoupleDefs2} \\
    $ G_o $ & Photon-phonon optomechanical coupling rate  &  \eqref{eq:GoDef} \\
    $ G_m $ & Phonon-phonon coupling rate between mechanical modes & \eqref{eq:GmDef} \\
    $ g_o $ & Convenient alternative definition of optomechanical coupling rate  &  \eqref{eq:gDef} \\
    $ \chi_c $ & Cavity susceptibility  &  \eqref{eq:CavSusc} \\
    $ R_b $ & Bright mode response function  &  \eqref{eq:Response} \\
    $ H_{j,0} $ & Hamiltonian for isolated harmonic oscillator with mass $m$ and eigenfrequency $\omega_j$  ($j = b,d$) &  \eqref{eq:H0b} \\
    $ x_{\mathrm{zpf},j,0}  $ & Position fluctuations in the ground state of $H_{j,0}$  ($j = b,d$) &  \eqref{eq:xzpf0} \\
    $ p_{\mathrm{zpf},j,0}  $ & Momentum fluctuations in the ground state of $H_{j,0}$  ($j = b,d$)&  \eqref{eq:xzpf0} \\
    $ x_{\mathrm{zpf},j}  $ & Position fluctuations in the ground state $|0\rangle$ of the basis in which $\hat{\rho}_j$ is thermal   ($j = b,d$)&  \eqref{eq:zpfDef1} , \eqref{eq:xzpf} , \eqref{eq:zpfWeak}\\
    $ p_{\mathrm{zpf},j}  $ & Momentum fluctuations in the ground state $|0\rangle$ of the basis in which $\hat{\rho}_j$ is thermal   ($j = b,d$)&  \eqref{eq:zpfDef2}  \eqref{eq:pzpf} \\
    $ \theta $ & Angle quantifying the symmetrized correlation between position and momentum & \eqref{eq:zpfDef1} , \eqref{eq:zpfDef2}  , \eqref{eq:anglesin}  \\
    $ \hat{b}_0  , \hat{d}_0 $ & Phonon annihilation operators for bright and dark modes  & \eqref{eq:PhononAnnihilationOperator} \\
    $ \hat{b} , \hat{d} $ & Annihilation operators for bright and dark modes for the number basis in which $\hat{\rho}_b , \hat{\rho}_d$ is thermal &  \eqref{eq:nmFromBs} , \eqref{eq:zpfDef1} , \eqref{eq:zpfDef2} , \eqref{eq:AnnihilationRelation} \\
    $ \bar{n}_{j,0} $ & Average phonon number ($j = b,d$)  &  \eqref{eq:PhononNumberDef} \\
    $ \bar{n}_b $ & Thermal occupation number in the basis in which $\hat{\rho}_b$ is thermal  &  \eqref{eq:nmFromBs} , \eqref{eq:BOccNum2} , \eqref{eq:nmWeak} , \eqref{eq:ThermOccStrongCoupling} , \eqref{eq:nmExact}  \\
    $ \mu_j $ & Quantum state purity of one-dimensional mode ($j = b,d$) &  \eqref{eq:Purity}, \eqref{eq:PurityThermal}  \\
    $  \psi_n  $ & Position space representation of the basis $|n\rangle$ in which $\hat{\rho}$ is thermal & \eqref{eq:Psin} \\
    $  M \Omega  $ & Parameter entering the wave functions $\psi_n$ & \eqref{eq:OmegaComplexTildelambdaDef} \\
    $\tilde{\omega}_b $ & Effective bright mode resonance frequency in the weak coupling regime & \eqref{eq:OptSpring} \\
    $\tilde{\gamma}_b $ & Effective bright mode decay rate in the weak couling regime & \eqref{eq:gammaTilde} \\
    $S_{x_b x_b} $ & Bright mode position spectral density   & \eqref{eq:SxxDef} \\
   $ \omega_\pm $ & Normal-mode (polariton) eigenfrequencies  &  \eqref{eq:PolFreqs1D}  \\
    $ \kappa_\pm $ & Normal-mode (polariton) decay rates  &  \eqref{eq:kappapm}  \\
    $ n_\pm $ & Average polariton numbers  &  \eqref{eq:npmStrong}  \\
    $ x_{\mathrm{zpf},\pm} $ & Bright mode position fluctuations when $n_\pm = 0$ & \eqref{eq:xzpfpmDef} \\
    $ p_{\mathrm{zpf},\pm} $ & Bright mode momentum fluctuations when $n_\pm = 0$ & \eqref{eq:pzpfpmDef} \\
    $ \mu_\mathrm{2D} $ & Quantum state purity of two-dimensional oscillator &  \eqref{eq:Purity2D} , \eqref{eq:Purity2DRWA_Max} , \eqref{eq:2D_QBAL1}-\eqref{eq:2D_QBAL5}   \\
    \end{tabular}
\end{ruledtabular}
\end{table}
\vspace{0.5cm}
\twocolumngrid


\end{document}